\documentstyle[aps,preprint]{revtex}  
\begin{document}
\draft
\tighten
\preprint{\vbox{Submitted to Physics Letters B 
                \hfill YUM 96-22\\
\null \hfill  SNUTP 96-104 }}
\title{QCD Sum Rule for  $\Lambda$(1405)}
\author{Hungchong Kim\footnote{E-mail:hung@phya.yonsei.ac.kr} and
Su Houng Lee\footnote{E-mail:suhoung@phya.yonsei.ac.kr}}
\address{Department of Physics, Yonsei University, Seoul, 120-749, Korea}
\maketitle
\begin{abstract}
Motivated by the recently constructed interpolation field for  
 S$_{11}$(1535),
we propose a new interpolating field for $\Lambda$(1405). Using this
current, we calculate the mass of $\Lambda$(1405) based on the conventional 
QCD sum rule analysis.  By calculating the Wilson coefficients up to 
dimension 8 operators and taking into account the mass corrections from 
s-quark, we find the calculated mass of $\Lambda$(1405) to be very close to 
its experimental value. 

\end{abstract}  
\pacs{PACS number(s):}

For many years, QCD sum rule  has been widely used to study the spectral 
properties of the hadrons since it was first introduced by 
Shifman, Vainshtein, and Zakharov~\cite{SVZ}.  The basic idea of the 
QCD sum rule is to describe the properties of a hadron, 
such as its mass or its 
coupling strength to other hadrons, in terms of QCD parameters.  
The starting point of the QCD sum rule is to introduce an appropriate
interpolating field for the hadron of concern using the quark and the 
gluon fields.
Successful description of the hadron depends partly on how strongly the 
interpolating field couples to the hadron.  Therefore, the specific form of
the interpolating field is usually determined by 
requiring it to have a  sufficiently large overlap with the 
hadron's nonrelativistic quark wave functions
as well as requiring the correct quantum numbers.  

Most QCD sum rules so far  focus on the lowest resonance while
putting all other higher resonances into the continuum~\cite{qsr}.  
However, often in nuclear physics, 
it is important to understand the properties of the  higher resonances and 
their decaying properties 
as they often participate in scattering processes. In this
regards, as a starting point of studying the excited baryonic states,
we have recently proposed a way to study the lowest excited state 
with negative-parity using the conventional QCD sum rule analysis~\cite{hung}. 
There, the interpolating field for S$_{11}$ is constructed by requiring it
to have a large overlap with the 
nonrelativistic quark wave function of S$_{11}$, as suggested by the
bag model~\cite{chodos} or the nonrelativistic quark model~\cite{isgur}.
The reasonable prediction of the S$_{11}$ mass and a stable Borel curve 
indicate
that our interpolating field introduced there seems to overlap strongly with
the S$_{11}$.   In this work, we will apply the same method to study  
the $\Lambda$ (1405).    ( In the following, we will denote $\Lambda$ (1405)
as $\Lambda^*$ and $\Lambda$ (1115) as $\Lambda$.)

First, we start by constructing the interpolating field for $\Lambda$.
$\Lambda$ has the quantum numbers of $I = 0$ and $S = -1$ composed
of (u, d, s) quarks.  The interpolating field for $\Lambda$  in 
most QCD sum rule analysis is obtained from the {\it Ioffe current} of 
nucleon using SU(3) transformation. 
Here, we directly construct the $\Lambda$ 
interpolating field following the procedure that has been used in 
constructing the nucleon interpolating field~\cite{griegel}.  

A simple way to construct the $\Lambda$ interpolating field is to
combine an up  and a down quark to a spin 0 isoscalar diquark and then 
attaching an s-quark to the diquark.  The quantum numbers of $\Lambda$ are
carried by the attached s-quark.  So the possible forms of $\Lambda$ 
interpolating field are 
\begin{eqnarray}
\eta_1 = \epsilon_{abc} (u^{\rm T}_a C \gamma_5 d_b) 
s_c\;; \quad \eta_2 = \epsilon_{abc} (u^{\rm T}_a C d_b) 
\gamma_5 s_c\ .
\end{eqnarray}
In general, one can choose the interpolating field of $\Lambda$ to be an 
arbitrary linear combination of these two fields,
\begin{eqnarray}
\eta_{\Lambda} (t)  = 2 (t\ \eta_1 + \eta_2)\ .
\end{eqnarray} 
This is one possible choice for the interpolating field with
the known quantum numbers of $\Lambda$.  Note that in the 
non-relativistic limit, $\eta_1$ reduces to three quarks in the 
s-wave state. 

The interpolating field for $\Lambda^*$ can be constructed by applying
$z \cdot D$ to one of the quark fields above. The four vector $z^\mu$ is 
orthogonal to the momentum of $\Lambda^*$, $q^\mu$ ($z \cdot q = 0 $) 
and $z^2 =-1$ so that, in the rest frame of $\Lambda^*$, 
$z \cdot D$ reduces to the derivative in the space direction.  
By introducing the covariant derivative in this way,  
the interpolating field will pick out one quark in the p-wave state and two
quarks in the s-wave state, which is believed to be the quark configuration 
of $\Lambda^*$ in the 
bag model~\cite{chodos} or nonrelativistic quark model~\cite{isgur}.  
From the three possible choices of putting in the derivative,
we choose the following interpolating field,
\begin{eqnarray}
\eta_{\Lambda^*} (t) = 2 \epsilon_{abc} 
                     [ t\ (u^{\rm T}_a C \gamma_5 z \cdot D d_b) s_c\
                     +  (u^{\rm T}_a C z \cdot D d_b) \gamma_5 s_c ]\ .
\end{eqnarray}
This choice is preferred because the correlator calculated from this field
has the nonzero contribution from  the lowest order chiral
breaking term, $\langle {\bar q} q \rangle$.  One other choice with
the covariant derivative acting on the u-quark yields the same expression for
the correlator due the isospin symmetry.  When the derivative acts on 
the s-quark, the
correlator has zero contribution from $\langle {\bar q} q \rangle$.
The parameter $t$ will be determined by optimization procedure 
introduced in Ref.~\cite{hung}.

With this current, we consider the following time-ordered correlation function,
\begin{eqnarray}
\Pi(q) = \int dx^4 e^{iq\cdot x} i\ \langle 0 | {\rm T} [ \eta_{\Lambda^*} (x)
{\bar \eta_{\Lambda^*}} (0)] |
0 \rangle\ ,
\label{cor}
\end{eqnarray}
which will have two invariant scalar functions defined as,
\begin{equation}
\Pi(q) = \Pi_1(q^2, z^2) + \Pi_q(q^2,z^2) \not\!q \ .
\label{pi}
\end{equation}  

The so called {\it theoretical side} of the sum rule is obtained by 
calculating the two scalar functions in the operator product expansion 
(OPE) whose Wilson coefficients  are calculated using
the fix-point gauge and the standard background-field techniques~\cite{nov}.
Our calculation up to dimension 8 operators gives
\begin{eqnarray}
\Pi^{\rm ope}_q (q) &=& - { q^6 {\rm ln} (-q^2) \over 2^{6}\times 3^2 \times
                         5\times \pi^4 } ( t^2 + 1)
                        - { q^2 {\rm ln} (-q^2) \over 2^{9}\times
                         9 \times \pi^2 } (9 t^2 -2 t + 9)
                        \langle \frac{\alpha_s}{\pi} {\cal G}^2
                        \rangle \nonumber \\
                    &&- { t^2 -1 \over 12 q^2} \langle {\bar q} q \rangle
                        \langle g_s {\bar q} \sigma \cdot {\cal G} q \rangle
                      -{ q^2 {\rm ln} (-q^2) \over 60 \pi^2 } 
                       ( t^2 + 1) m_s\langle {\bar q} q \rangle \nonumber\ \\
                     && +{(19t^2+4t+19) {\rm ln} (-q^2) \over 3^2 
                       \times 2^6 \times 5 \times \pi^2}
                       m_s \langle g_s {\bar q} \sigma \cdot {\cal G} q 
                       \rangle\
                      - {t^2+1 \over  240} { m_s \over q^2} 
                       \langle \frac{\alpha_s}{\pi} {\cal G}^2 \rangle
                         \langle {\bar q} q \rangle\label{ope1} \ ,\\
\Pi^{\rm ope}_{1} (q) &=& {q^4 {\rm ln} (-q^2) \over 60 \pi^2 }(t^2-1)
                          \langle {\bar q} q \rangle 
                     - { q^2 {\rm ln} (-q^2)
                     \over 60 \pi^2}
                     ( t^2  -1)
                    \langle g_s {\bar q} \sigma \cdot {\cal G} q \rangle
\nonumber \\
                    && + { {\rm ln} (-q^2) \over 120}
                         (t^2-1)
                         \langle \frac{\alpha_s}{\pi} {\cal G}^2 \rangle
                         \langle {\bar q} q \rangle\
                       -{ q^6 {\rm ln} (-q^2) \over 2^{7} \times
                         5\times \pi^4 } m_s ( t^2 - 1)
\nonumber \\
                      && -{ q^2 {\rm ln} (-q^2) \over 2^{6}\times
                         3 \times \pi^2 } ( t^2 - 1)
                        m_s \langle \frac{\alpha_s}{\pi} {\cal G}^2
                        \rangle  .
\label{ope2}
\end{eqnarray}
Note that, in calculating these, we
have made use of the properties, $z\cdot q =0$ and $z^2 =-1$.
Note also that we have calculated the mass correction ($m_s$) from s-quark
up to dimension 8.  The mass corrections come either from the s-quark
propagator or from the short distant expansion of the condensate involving
s-quark. 
As has been noted in Ref.~\cite{hung}, the introduction of the covariant
derivative enhances the contribution from the dimension five operator,
$\langle g_s {\bar q} \sigma \cdot {\cal G} q \rangle\ $, in 
 $\Pi^{\rm ope}_1$ and constitutes
the major part of the mass splitting between the positive-parity ground state 
baryon and the negative-parity first excited state.
In our case, this condensate also contributes to $\Pi^{\rm ope}_{q}$  through
s-quark mass. 
As in  Ref.~\cite{hung}, we did not include the contribution
from the dimension six operator,  $\langle {\cal G}^3 \rangle$, as it is
of order $\alpha^{3/2}$.   

Due to the orthogonal property of $z_\mu$ ( $z \cdot q=0$), the 
{\it phenomenological side} of the correlator  
take the following form,
\begin{eqnarray}
\Pi^{\rm phen} (q) &=&  - \Bigr [ \lambda_{\Lambda}^2 {  \not\!q - 
               M_{\Lambda} \over
                  q^2 -M_{\Lambda}^2 +i\epsilon }
                  + \lambda_{\Lambda^*}^2 {\not\!q + M_{\Lambda^*} \over
                 q^2 -M_{\Lambda^*}^2 + i\epsilon } \Bigl ]+
                 {\rm continuum}\label{phe}\  \\
                 &\equiv& \Pi^{\rm phen}_1 +\not\!q \Pi^{\rm phen}_q\ .
\label{phe1}  
\end{eqnarray}
The lowest state with positive ( negative) parity
is identified with the $\Lambda$ ($\Lambda^*$) where  
$\lambda_{\Lambda}$ ($\lambda_{\Lambda^*}$) denotes its 
coupling strengths to our interpolating field. 
We have put all other resonances 
with higher masses into two separate continuum, one for the 
positive-parity resonances starting at  $s_+$, the other for the 
negative-parity resonances starting at  $s_-$. 
The form of these continuum parts of the phenomenological side are 
obtained as follows.  First, look at the corresponding continuum part of the 
imaginary part of the OPE.  Second
use dispersion relation to obtain the real part but assuming that the cut
starts at the continuum threshold $s_+$ or $s_-$.

The parameter $t$ is chosen in such a way that the resulting interpolating
field does not couple to the positive-parity baryon while maximally couples
to the negative-parity baryon.  For this, we follow  Ref.~\cite{hung,oka}.  
There we introduced the two scalar functions, $\Pi_1^o$ and $\Pi_q^o$, at the
rest frame (${\bf q} =0$), which are the two invariant functions in the 
 ``old-fashioned'' correlation function $\Pi^o(q)= i \int d^4x e^{iqx} 
\theta(x_0) \langle 0| \eta(x) \eta(0) |0 \rangle$~\cite{oka}.  
In the sum (difference) of the
two scalar functions, only the positive (negative) parity states contributes 
so that one can construct a finite energy sum rule which gives the coupling 
strength $\lambda_{\Lambda}^2$ as a function of $t$,
\begin{eqnarray}
60 \pi^2 \lambda_{\Lambda}^2 &=& { t^2 +1 \over 48 \pi^2} 
{ (\sqrt {s_+})^8 \over 8} 
+ { 5 \over 2^7 \times 3} (9t^2 -2 t + 9)
  \langle {\alpha_s \over \pi} 
  {\cal G}^2 \rangle {(\sqrt{s_+})^4 \over 4}\nonumber \\ 
&+&{5 \pi^2 \over 2} (t^2 -1) \langle {\bar q} q \rangle 
  \langle g_s {\bar q} \sigma \cdot {\cal G} q \rangle 
   + (t^2 + 1) m_s \langle {\bar q} q \rangle {(\sqrt{s_+})^4 \over 4}
\nonumber \\
&-& {19 t^2 + 4 t + 19 \over 48} 
m_s \langle g_s {\bar q} \sigma \cdot {\cal G} q \rangle 
{(\sqrt{s_+})^2 \over 2} + {\pi^2 \over 8} (t^2+1) 
m_s \langle {\bar q} q \rangle \langle {\alpha_s \over \pi}{\cal G}^2 \rangle
\nonumber \\
&+& (t^2-1)\langle {\bar q} q \rangle {(\sqrt{s_+})^5 \over 5} 
- (t^2-1) \langle g_s {\bar q} \sigma \cdot {\cal G} q 
\rangle {(\sqrt{s_+})^3 \over 3} 
\nonumber \\
&+& {\pi^2 \over 2} (t^2 -1)\sqrt{s_+} \langle {\bar q} q \rangle 
\langle {\alpha_s \over \pi}{\cal G}^2 \rangle - 
{3 (t^2 -1) \over 32 \pi^2} {(\sqrt{s_+})^7 \over 7}
\nonumber \\
&-& {5 \over 16} (t^2 -1) m_s \langle {\alpha_s \over \pi}{\cal G}^2 \rangle
{(\sqrt{s_+})^3 \over 3}\ .
\end{eqnarray}
The possible $t$ is chosen so that $\lambda_{\Lambda}^2$ is zero.
Normally there are two $t$s which give $\lambda_{\Lambda}^2=0$.  
We take $t$ which yields larger value of $\lambda_{\Lambda^*}^2$.

Now we are in a position to determine the mass of $\Lambda^*$. 
We equate the phenomenological side of the two invariant functions in 
Eq.~(\ref{phe1}) 
to their corresponding OPE sides in Eq.~(\ref{ope1}) and Eq.~(\ref{ope2}).  
After taking the 
Borel transformation and taking the ratio of the two sum rules, we obtain a
Borel sum rule for the  $M_{\Lambda^*}$ which yields,

\begin{eqnarray}
M_{\Lambda^*} &=& \Big[ -[\Delta^6_- + \Delta^6_+] 
(t^2-1)\langle {\bar q} q \rangle +
{\Delta^4_-+ \Delta^4_+ \over 2}(t^2-1)\langle
 g_s {\bar q} \sigma \cdot {\cal G}
                         q \rangle\  \nonumber\ \\
     && -{\Delta^2_-+ \Delta^2_+ \over 4} \pi^2 (t^2-1)
                 \langle {\alpha_s \over \pi} {\cal G}^2 \rangle
                 \langle {\bar q} q \rangle 
            +{\Delta^8_- + \Delta^8_+ \over 32 \pi^2}
            (t^2-1) 9 m_s \nonumber\ \\
          &&   + {\Delta^4_- + \Delta^4_+ \over 32} 5
             (t^2 -1) m_s \langle {\alpha_s \over \pi} {\cal G}^2 \rangle
\Big]  \nonumber \\
&\times& \Big[ { \Delta^8_-+ \Delta^8_+ \over 16 \pi^2} ( t^2 +1) 
       +{\Delta^4_- + \Delta^4_+ \over 2^8 \times 3}\langle
  {\alpha_s \over \pi} {\cal G}^2 \rangle 5 (9 t^2 -2t + 9)  \nonumber\ \\
    && + 5\pi^2 (t^2-1) \langle {\bar q} q \rangle
         \langle g_s {\bar q} \sigma \cdot {\cal G} q \rangle\ 
        + {\Delta^4_-+ \Delta^4_+ \over 2}
        m_s \langle {\bar q} q \rangle (t^2 + 1)
\nonumber \\
       && - {\Delta^2_-+ \Delta^2_+ \over 96} m_s
        \langle g_s {\bar q} \sigma \cdot {\cal G} q \rangle\ 
        (19 t^2 +4t+19) \nonumber \\
       && + {\pi^2 \over 4}m_s\langle {\bar q} q \rangle\langle
          {\alpha_s \over \pi} {\cal G}^2 \rangle (t^2 + 1) \Big]^{-1},
\label{mass}
\end{eqnarray}
where we have defined
\begin{eqnarray}
\Delta^2_{\pm} &=& M^2(1 -e^{-s_{\pm}/M^2})\ ,\\
\Delta^4_{\pm} &=& M^4 \Bigl [1 -
(1 + {s_{\pm} \over M^2})e^{-s_{\pm} /M^2} \Bigr ] 
\ ,\\
\Delta^6_{\pm} &=& M^6 \Bigl [1 - (1 + {s_{\pm} \over M^2} +
{s_{\pm}^2 \over 2 M^4})e^{-s_{\pm} /M^2} \Bigr ] \ ,\\
\Delta^8_{\pm} &=& M^8 \Bigl [1 - ({s_{\pm} \over M^2}
+{s_{\pm}^2 \over 2 M^4}+{s_{\pm}^3 \over 6 M^6}) e^{-s_{\pm}/M^2}\Bigr ] \ .
\end{eqnarray}

We take the following QCD parameters which are conventionally found in
literature, 
\begin{eqnarray}
\langle {\bar q} q \rangle = -(0.23\ {\rm GeV})^3\;; \quad 
\langle {\alpha_s \over \pi} {\cal G}^2 \rangle = (0.35\ {\rm GeV})^4\;; \quad 
m_s = 150~~{\rm MeV}\ .
\end{eqnarray}
However, our final result is not sensitive to these
parameters as long as they are varied within  known error bars.  
The continuum thresholds are chosen as the positions of the 
next higher resonances:
\begin{eqnarray}
s_+ = (1.6)^2~~{\rm GeV}^2\;; \quad s_-=  (1.67)^2~~{\rm GeV}^2\ .
\end{eqnarray}
The most important parameter for our prediction is the quark-gluon
condensate which is usually expressed in terms of the quark condensate
$\langle {\bar q} q \rangle$:
\begin{equation}
\langle g_s {\bar q} \sigma \cdot {\cal G} q \rangle\ \equiv
2\lambda^2_q \langle {\bar q} q \rangle\ .
\label{qgc}
\end{equation}
The value of $\lambda_q^2$, which represents the average virtual momentum of 
vacuum quarks, is  not well known.  In our previous work on
S$_{11}$~\cite{hung}, we have found that the best prediction on S$_{11}$ was
obtained for $\lambda_q^2 = 0.435$ GeV$^2$.   So we take this value.

\begin{figure}
\caption{Our prediction for $\Lambda^*$ mass versus Borel mass.}
\centerline{%
\vbox to 2.4in{\vss
   \hbox to 3.3in{\includegraphics{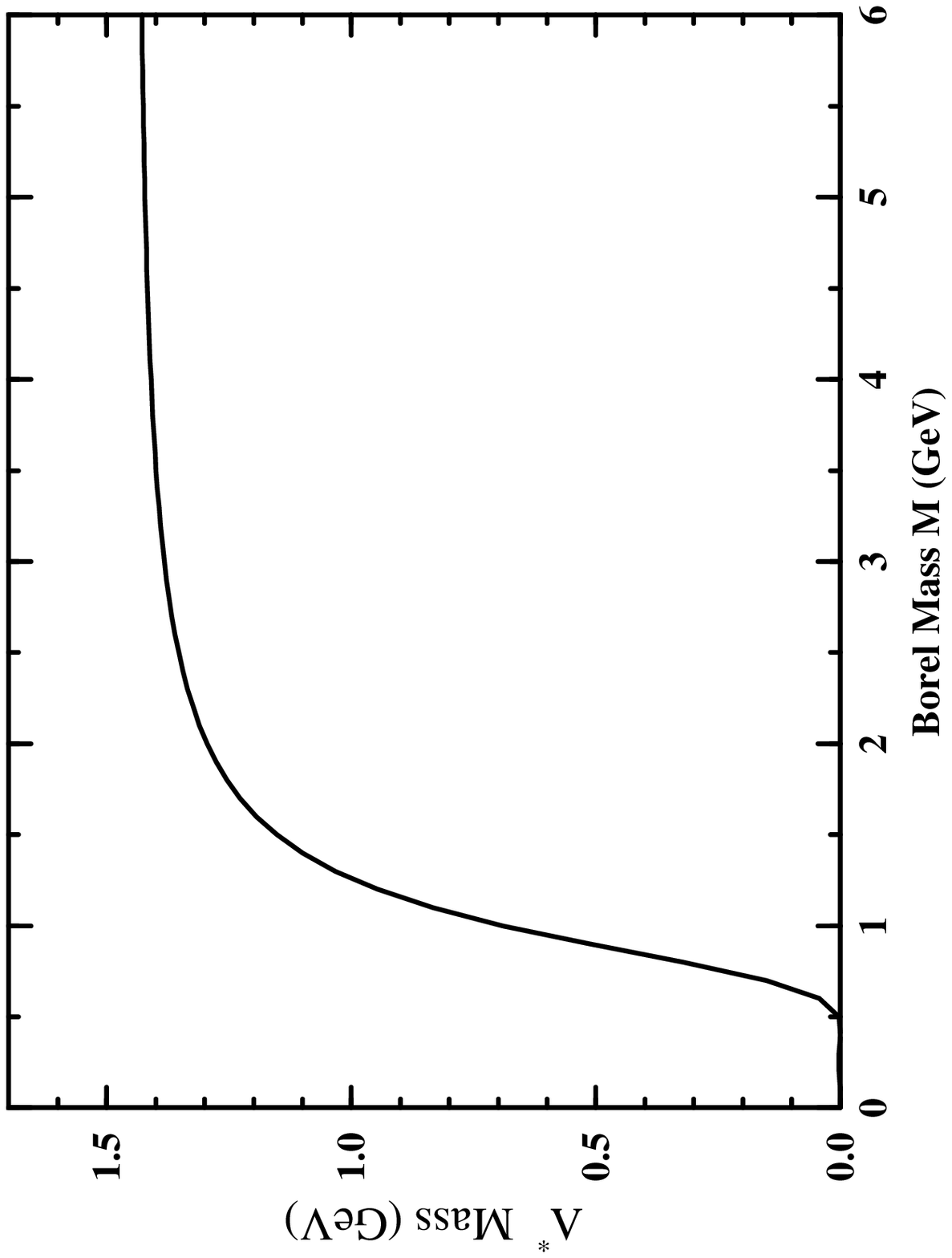}\hss}}
}
\bigskip
\vspace{400pt}

\label{fig}
\end{figure}

With these QCD parameters, the optimal $t$ is found to be $-2.781$.
Putting all these into Eq.~(\ref{mass}), we obtain the Borel curve shown 
in Fig.~\ref{fig}.  
Clearly we have a Borel plateau as the Borel mass increases. 
Our prediction  for $\Lambda^*$ mass which is obtained from
most stable Borel plateau is  $M_{\Lambda^*} =1.42$ GeV.  This agrees
with its experimental value of 1.405 GeV within 2 \%.

In summary, we proposed a new interpolating field for $\Lambda^*$(1405)
constructed to reproduce its quark configuration in the bag model or
the nonrelativistic quark model.  By calculating the Wilson coefficients
up to dimension 8 operators in the conventional QCD sum rule approach,  
we found that our interpolating field yields the mass of $\Lambda^*$
close to its known value of 1405 MeV within 2 \% .  Therefore, our
interpolating field couples strongly to $\Lambda^*$ and can be
used to study its other spectral properties.

\acknowledgments
SHL was supported in part by the by the Basic Science Research
Institute program of the Korean Ministry of Education through grant no.
BSRI-96-2425, by KOSEF through the CTP at Seoul National University
and by Yonsei University Research Grant. 
HK  was supported by KOSEF.  We thank D. K. Griegel and S. Choe for useful
discussion.

\end{document}